\documentclass[referee,useAMS,usenatbib,usegraphicx]{mn2e}
\newcommand{\Msun}{\mbox{$M_{\odot}$}}
\newcommand{\Rsun}{\mbox{$R_{\odot}$}}

\newcommand{\kms}{\mbox{km s$^{-1}$}}

\title[NSVS\,06507557; a low-mass double-lined eclipsing binary ]
      {NSVS\,06507557; a low-mass double-lined eclipsing binary \thanks{Based on photometric and spectroscopic observations collected at T\"{U}B\.{I}TAK National Observatory (Turkey).}\thanks{Table 1 is only avaliable in electronic form at the CDS via ftp to http://www.blackwell-syngery.com/doi....}}
\author[]
  {\"{O}. \c{C}ak{\i}rl{\i}$^1$\thanks{e-mail:omur.cakirli@ege.edu.tr} and C.~\.{I}bano\v{g}lu$^1$\\ \\
  $^1$Ege University, Science Faculty, Astronomy and Space Sciences Dept., 35100 Bornova, \.{I}zmir, Turkey\\
  }

\date{Released 2009 Xxxxx XX}

\pagerange{\pageref{firstpage}--\pageref{lastpage}} \pubyear{2008}
\begin{document}
\maketitle
\label{firstpage}
\begin{abstract}
In this paper we present the results of a detailed spectroscopic and photometric analysis of the V=13$^m$.4 low-mass 
eclipsing binary NSVS\,06507557 with an orbital period of 0.515 d. We obtained a series of mid-resolution spectra covering nearly entire orbit of the 
system. In addition we obtained simultaneous VRI broadband photometry using a small aperture telescope. From these 
spectroscopic and photometric data we have derived the system's orbital parameters and determined the fundamental 
stellar parameters of the two components. Our results indicate that NSVS\,06507557 consists of a K9 and an M3 
pre-main-sequence stars with masses of 0.66$\pm$0.09 \Msun and 0.28$\pm$0.05 \Msun and radii of 0.60$\pm$0.03 
and 0.44$\pm$0.02 \Rsun, located at a distance of 111$\pm$9 pc. The radius of the less massive secondary component is 
larger than that of the zero-age main-sequnce star having the same mass.
While the radius of the primary component is in agreement with ZAMS the secondary component appers to be larger by about 35 \%
with respect to its ZAMS counterpart. Night-to-night intrinsic light 
variations up to  0$^m$.2 have been observed. In addition, the H$_{\alpha}$, H$_{\beta}$ lines and the 
forbidden line of [O{\sc i}] are seen in emission. The Li{\sc i} 6708 \AA~ absorption line is seen in most 
of the spectra. These features are taken to be the signs of the classic T Tauri stars' characteristics.  The 
parameters we derived are consistent with an age of about 20 Myr according to the stellar evolutionary models. The 
spectroscopic and photometric results are in agreement with those obtained by theoretical predictions.
\end{abstract}


\section{introduction}
Low-mass stars constitute the majority of stars by number in our Galaxy. Since their lower masses with respect 
to the Sun they have also very low intrinsic brightness. Although the intrinsic faintness of these stars many 
low-luminosity stars were discovered particularly by the near-infrared sky surveys: Deep Near Infrared Survey 
(Delfosse et al. 1997), Two Micron All Sky Survey (Skrutskie et al. 1997), Sloan Digital Sky Survey (York et al. 
2000), Northern Sky Variability Survey (Wozniak et al. 2004). Since their 
main-sequence lifetimes are considerably longer than the age of the universe both the young and old low-mass 
stars are located on the lower right part of the the HR diagram. Low-mass stars surround many important regions 
of stellar parameter space which include the onset of complete convection in the stellar interior, the onset of 
electron degeneracy in the core, and the formation of dust and depletion metals onto dust grains in the stellar 
atmosphere (West et al. 2004).  Recent studies have shown that while the observed radii of the low-mass stars 
are significantly larger than those predicted by current stellar models, in contrast their effective temperatures 
are cooler (Ribas et al. 2008, Lopez-Morales and Ribas 2005). Chabrier, Gallardo \& Baraffe (2007) have put
forward the hypothesis that the observed radius and temperature discrepancies are consequences of the convection due to rotation and/or 
magnetic field and the presence of large surface magnetic spots. Therefore low-mass stars are key interest in 
studies of both formation of the stars in star-forming regions and comparison their parameters with those 
predicted from theoretical stellar models.  
 
The fundamental parameters such as mass, radius, effective temperature and luminosity, all in a distance-independent manner, of a star could be determined empirically from eclipsing binary stars.  Precise masses and radii can be determined from multi-wavelength photometry and spectroscopy, obtained with current technology, of double-lined close binary systems. However, the number of well-studied eclipsing binaries with low-mass components is rather small because of their low intrinsic brightness. Furthermore, most of their  light curves are undergone strong distortion due to magnetic activity. Therefore, multi-passband photometric and spectroscopic observations of additional low-mass binaries would be extremely useful.

The binary nature of the star known as NSVS 06507557 (=2MASS J01582387+2521196, hereafter NSVS 0650) was discovered by Shaw and Lopez-Morales (2006) using the database NSVS (Wozniak et al. 2004). The eclipse period was determined to be 0.515 d. Later on the first VRI light curves and preliminary models are presented by Coughlin and Shaw (2007). Taking the BVRI magnitudes from the USNO NOMAD catalog and JHK from 2MASS catalog they estimated the effective temperature of 3860 K for the primary, corresponding to a spectral type of M0V.  As they have noted a difficulty encountered in modeling was the high-level spot activity of the components. Not only the radii and effective temperatures of the component stars were determined but also the rough masses estimated by them.  We have conducted a photometric and spectroscopic monitoring program of several low-mass eclipsing binaries. In this paper we present, the results of multi-wavelength optical photometry and spectroscopy for double-lined eclipsing binary NSVS 0650.

\section{observation}
\subsection{Photometry}
NSVS\,0650757 was first identified in the {\sc Northern Sky Variability Survey} (NSVS; Wozniak et al. 2004) as a detached 
eclipsing binary system with a maximum, out-of-eclipse V-bandpass magnitude $V$=13$^m$.05 and a period of P=0.51509 day. The data 
from the NSVS, obtained with the Robotic Optical Transient Search Experiment telescopes (ROTSE), contains positions, light curves 
and V magnitudes for about 14 million objects ranging in magnitudes from 8 to 15.5. The B, V, R, and I magnitudes for NSVS 0650 
were  listed in the USNO NOMAD catalog as ( {\sc Naval Observatory Merged Astronomical Dataset}, NOMAD-1.0, Zacharias et 
al. 2004), $B$=14$^m$.53, $V$=13$^m$.37, $R$=12$^m$.47; on the other hand the infra-red magnitudes in three bandpasses were 
given as  $J$=10$^m$.918 $H$=10$^m$.267, and $K$=10$^m$.092 in the 2MASS catalog (Cutri et al. 2003).

In the NSVS survey, 262 V-bandpass measurements of the variable were obtained during the period June 1999 - March 2000 with a 
median sampling rate of 0.25$^{-1}$. The resulting light curve exhibits periodic eclipses with a depth of $\sim$0$^m$.7 in the 
deeper eclipse and the mean standard deviation in the out-of-eclipse phases was about 0$^m$.073.

The photometric observations of NSVS 0650 were carried out with the 0.4 m telescope at the Ege University Observatory. The 
0.4 m telescope equipped with an Apogee 1kx1k CCD camera and standard Bessel VRI bandpasses. The observations were performed on 
seven nights between September 01 and November 30, 2008. To get the higher accuracy the target NSVS0650 was placed near to the 
center of the CCD and three nearby stars located on the same frame were taken for comparison. The stars GSC 01760-01860 and USNO A2.0 
1125~638990 were selected as comparison and check, respectively. Therefore the target and comparison stars could be observed 
simultaneously with an exposure time of 10 seconds. Since the variable is very cool, red star the signal-to-noise ratio was highest 
in I- and lowest in the V-bandpass.  The differential observations of the comparison stars showed that they are stable during time 
span of our observations.  The data were processed with standard data reduction procedures including bias and over scan 
subtraction, flat-fielding, and aperture photometry. A total of 743, 812 and 612 photometric measurements were obtained in each V, R and I 
bandpasses, respectively. The average uncertainity of each differential measurement was less than 0$^m$.030. The V-, R- and I-bandpass  
magnitude differences, in the sense of variable minus comparison, are listed in Table 1 (available in the electronic form at the CDS).

\begin{table*}
\caption{Differential photometric measurements of NSVS\,0650 in the V, R and I bandpasses.}
\begin{tabular}{|c|c|c|c|c|c}
\hline
HJD(2\,400\,000+)  & $\Delta$V & HJD(2\,400\,000+) & $\Delta$R & HJD(2\,400\,000+) & $\Delta$I \\
\hline
54725.35259		&1.6025		& 54725.35296	 & 1.1780  & 54725.35326 & 0.5799  \\
54725.35370		&1.6304	 	& 54725.35406	 & 1.1572  & 54725.35437 & 0.5563  \\
...	&...	&... & ...&...	&...	\\
...	&...	&... & ...&...	&...	\\

\hline
\end{tabular}
\end{table*}

The light curve shows a deep primary eclipse with an amount of 0$^m$.70 in the V-bandpass  and a shallow secondary eclipse with 
an amount of 0$^m$.23 which are clearly separated in phase, as is typical of fully detached binaries. The  primary and secondary 
eclipses occur almost 0.5 phase interval, indicating nearly circular orbit. An inspection of the nightly light curves presented in Fig. 1
clearly indicates considerable out-of-eclipse light variations up to 0$^m$.2.  This intrinsic variation of the binary system 
manifests itself in the deeper primary eclipse. 

\begin{figure*}
\includegraphics[width=16cm]{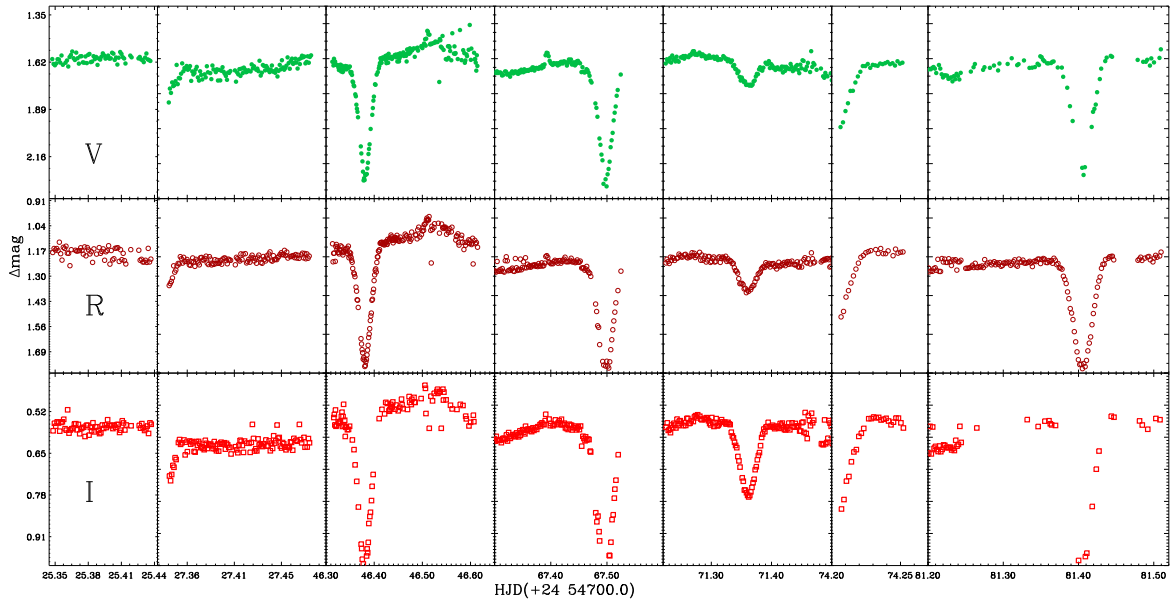}
\caption{The V-, R-, and I-bandpass nightly light curves for NSVS\,0650 from top to bottom. The
V-, R-, and I-bandpass light curves clearly show that the brightness of the variable significantly 
varies from night to night, particularly in out of eclipse.}
\label{Fig. 1.}
\end{figure*}

\subsection{Orbital period and ephemeris}
The first orbital period for 
NSVS\,0650 was determined as P=0.51509957 d by Shaw and Lopez-Morales (2006) from the NSVS database. Later on Coughlin \& Shaw (2007) 
observed seven low-mass detached systems, including NSVS\,0650, with the Southeastern Association for Research in Astronomy (SARA) 
0.9 m telescope. An orbital period of P=0.5150895$\pm$0.0000008 days, and an initial epoch T$_0$(HJD)=2453312.3722$\pm$0.0005 for 
the mid-primary eclipse were calculated using a least square fit. Partial primary and secondary eclipses which were detected in the 
time series photometric data were used in combination with the NSVS photometry to derive this ephemeris for the system. 

We obtained three times of mid-primary and one secondary eclipse during our observing run. The mid-eclipse timings and their 
standard deviations are caculated using the method of Kwee \& van Woerden (1956). These timings of the eclipses were 
listed in Table 1 together with two primary and a secondary eclipse collected from literature. The times for mid-eclipses 
are the average of times obtained in three bandpasses. We  define the epoch of the system, T$_0$, to be the midpoint of the 
most complete primary eclipse. For this reason we use the V-, R-, and I-bandpass data obtained on JD=2\,454\, 746 which 
cover almost the whole primary eclipse. A linear least square fit to the data listed in Table 1 yields the new ephemeris as,
 \begin{equation}
Min\, I\, (HJD)= 24\,54746.3801(5)+0.51508836(9)\times E, 
\end{equation}  
where E corresponds to the cycle number.  The residuals in the last column of Table 2 are computed with the new ephemeris. While 
the orbital period is nearly the same with that determined by  Coughlin \& Shaw (2007) its uncertainty is now very smaller than 
estimated by them. In the computation of the orbital phase for individual observations we used this ephemeris.

\begin{table}
\caption{Times of minima measured from the $VRI$-bandpass light curves.}
\begin{tabular}{|c|c|c|c}
\hline
HJD(2\,400\,000+)  & E	&Type          &O-C \\
\hline
51537.1282** 		&-6230.5  & II  &  0.0099  \\
51581.1569** &-6145.0   & I   & -0.0015  \\
53312.3722$^{\dagger}$  &-2784.0  & I   & -0.0006  \\
54746.3809$\pm$0.0005   		&0.0      & I   &  0.0000  \\
54767.4978$\pm$0.0006   		&41.0     & I   & -0.0018  \\
54771.3631$\pm$0.0004   		&48.5     & II  &  0.0004  \\
54781.4056$\pm$0.0002   		&68.0     & I   & -0.0014  \\
\hline
\end{tabular}
\begin{list}{}{}
\item[$^{**}$]{\small From the NSVS database.} 
\item[$^{\dagger}$]{\small Coughlin \& Shaw (2007).} 
\end{list}
\end{table}

\subsection{Intrinsic light variations}
The light curve of  NSVS 0650 shows two well-separated eclipses, as a typical of detached eclipsing binaries. The phase difference 
between the eclipses is about 0.5 which indicates a nearly circular orbit. Since the depths of the eclipses are very different, 
indicating that the components have unequal effective temperatures. The light variation both in primary and out-of-eclipse is 
clearly seen in all bandpasses. This light variation of about 0.2 mag peak-to-peak in the out-of-eclipse portions of the light 
curve reveals that there is an intrinsic variation in one or both components of the system. The amplitude of the intrinsic 
variations seems to larger with longer wavelengths. The light variations observed on JD~2454746 with long duration, just between 
primary and secondary eclipses, and also on JD~2454767 with very short duration, resemble a flare-like event which is common 
in M-type dwarf stars.

The data obtained by us are concentrated on seven nights ranging a time span of 56 days. The stars having masses smaller than 
that of the Sun are known to be heavely spotted. Therefore the out-of-eclipse light variations may be attributed to large 
spots on the surface of one or both component stars. In addition, flares on the less massive star cannot be ignored. However, it 
should be noted that the intrinsic light variations do not resemble to those observed in the spotted stars. A spot or spot 
groups on one or both components produces usually wave-like distortion on their light curves.  However, the out-of-eclipse 
light variations in NSVS 0650 seem to not correlated with the orbital period. 

\subsection{Spectroscopy}
Optical spectroscopic observations of NSVS\,0650 were obtained with the Turkish Faint 
Object Spectrograph Camera (TFOSC) attached to the 1.5 m telescope on 3 nights (September 15, 16, and 17, 2008) 
under good seeing conditions. Further details on the telescope and the spectrograph can be found at 
http://www.tug.tubitak.gov.tr. The wavelength coverage of each spectrum was 
4100-8100 \AA~in 11 orders, with a resolving power of $\lambda$/$\Delta \lambda$ 7\,000 at 6563 \AA~and an average 
signal-to-noise ratio (S/N) was $\sim$120. We also obtained a high S/N spectrum of the M dwarf GJ\,740 (M0 V) and GJ\,623 (M1.5 V) 
for use as templates in derivation of the radial velocities (Nidever et al. 2002). 

The electronic bias was removed from each image and we used the 'crreject' option for cosmic ray removal. Thus, the resulting 
spectra were largely cleaned from the cosmic rays. The echelle spectra were extracted and wavelength calibrated by
using Fe-Ar lamp source with help of the IRAF {\sc echelle} package. 
 
The stability of the instrument was checked by cross correlating the spectra of the standard star against each other using 
the {\sc fxcor} task in IRAF. The standard deviation of the differences between the velocities measured using {\sc fxcor} and the 
velocities in Nidever et al. (2002) was about 1.1 \kms.

\subsubsection{Spectral classification}
We have used our spectra to reveal the spectral type of the primary component of NSVS\,0650. For this purpose we have degraded the spectral
resolution from 7\,000 to 3\,000, by convolving them with a Gaussian kernel of the appropriate width, and we have measured
the equivalent widths ($EW$) of photospheric absorption lines for the spectral classification. We have followed the 
procedures of Hern\'andez et al. (2004), choosing helium lines in the blue-wavelength region, where the contribution 
of the secondary component to the observed spectrum is almost negligible. From several spectra we measured 
$EW_{\rm He I+ Fe I\lambda 4922 }=1.18\pm 0.12$\,\AA.

From the calibration relations $EW$--Spectral-Type of Hern\'andez et al. (2004), we have derived a spectral
type of K8 with an uncertainty of about 1 spectral subclass. The effective temperature deduced from the calibrations 
of Drilling \& Landolt (2000) or de Jager \& Nieuwenhuijzen (1987) is about 4\,050\,K. The spectral-type uncertainty 
leads to a temperature error of $\Delta T_{\rm eff} \approx 300$\,K.

The catalogs USNO, NOMAD and 2MASS provides BVRIJHK magnitudes for NSVS 0650. Using the observed colors of B-V=1.36$\pm$0.02 and V-I=2.13$\pm$0.02 mag 
and color-temperature relationships given by Drilling \& Landolt (2000) for the main sequence stars we estimate a spectral type K9$\pm$1 with an effective 
temperature of 3930$\pm$50 K for the primary star. The observed infrared colors of J-H=0.651$\pm$0.043 and H-K=0.175$\pm$0.038 given 
in the 2MASS catalog (Cutri et al. 2003) correspond to a spectral type of K9$\pm$2 is in a good agreement with that we derived   
by wide-band B-V and V-I photometric colors. We estimated a temperature of 3920$\pm$175 K from the calibrations of Tokunaga (2000). Temperature 
uncertainty of the primary component results from considerations of spectral type uncertainties, and calibration differences. The 
weighted mean of the effective temperature of the primary star is 3960$\pm$80 K. The effective temperature of the primary star what 
we derived from the photometric measurements is an a good agreement with that we estimated from the spectra alone. 

\begin{figure*}
\includegraphics[width=12cm]{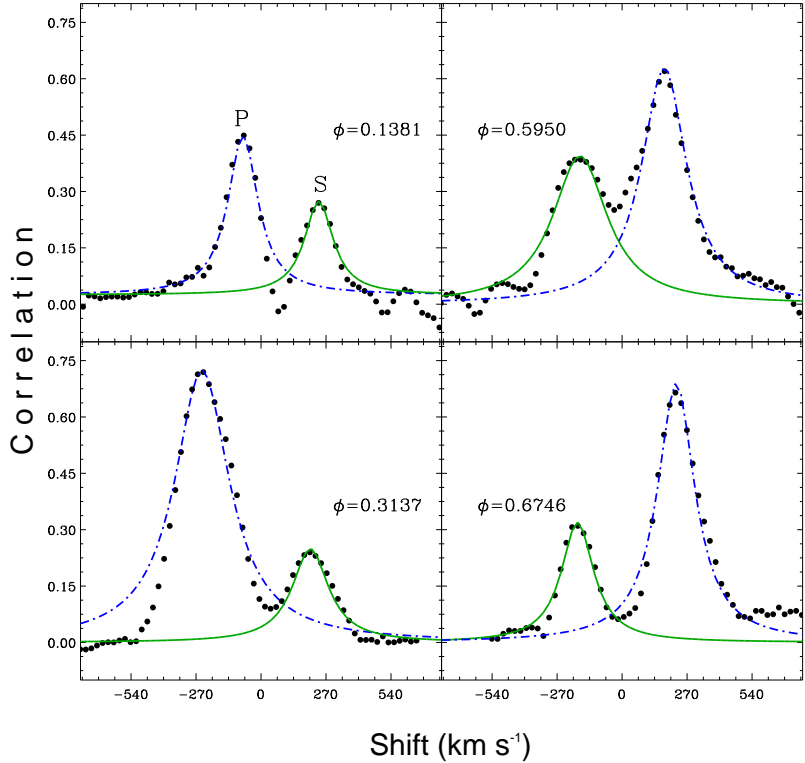}
\caption{Sample of Cross Correlation Functions (CCFs) between NSVS\,0650 and the radial velocity template 
spectrum around the first and second quadrature.}
\label{Figure 2.}
\end{figure*}

\begin{figure*}
\includegraphics[width=12cm]{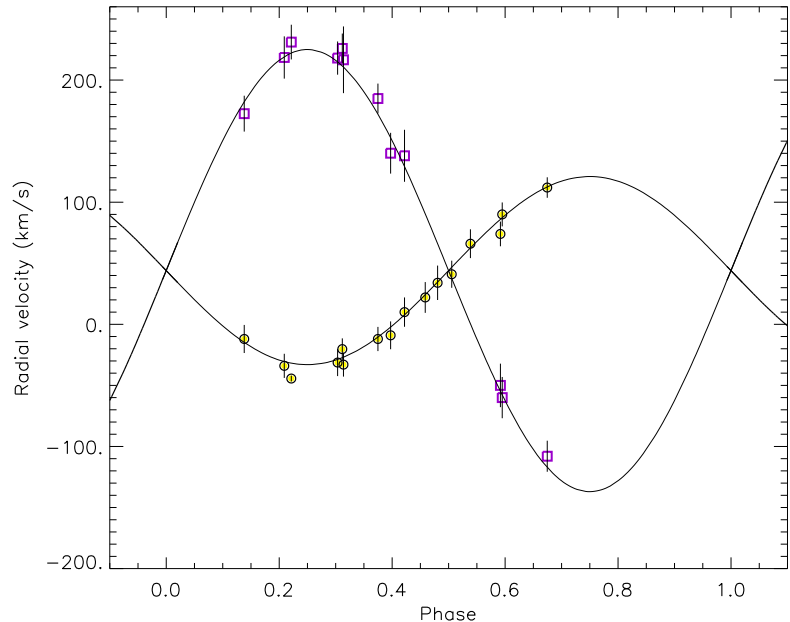}
\caption{Radial velocity curve folded on a period of 0.51508836 days, where phase zero is defined to 
be at primary mid-eclipse. Symbols with error bars show the RV measurements for two components 
of the system (primary: open circles, secondary: open squares). 
}
\label{Figure 3.}
\end{figure*}

\section{Analysis}
\subsection{Radial velocity curve}
To derive the radial velocities for the components of binary system, the 16 TFOSC spectraof the eclipsing binary  were cross-correlated against the spectrum of 
GJ\,740, a single-lined M0V star, on an order-by-order basis using the {\sc fxcor} package in IRAF. The majority of the spectra showed two 
distinct cross-correlation peaks in the quadrature, one for each component of the binary. Thus, both peaks were fit independently
in the quadrature with a Gaussian profile to measure the velocity and errors of the individual components. If the two peaks appear 
blended, a double Gaussian was applied to the combined profile using {\it de-blend} function in the task. For each of the 16 observations we 
then determined a weighted-average radial velocity for each star from all orders without significant contamination by telluric absorption
features. Here we used as weights the inverse of the variance of the radial velocity measurements in each order, as reported 
by {\sc fxcor}. In these data, we find no evidence for a third component, since the cross-correlation function showed only two distinct peaks.  

We adopted a two-Gaussian fit algorithm to resolve cross-correlation peaks near the first and second quadratures when 
spectral lines are visible separately. Figure\,2 shows examples of cros-correlations obtained by using the largest 
FWHM at nearly first and second quadratures. The two peaks, non-blended, correspond to each component of NSVS\,0650. The stronger 
peaks in each CCF correspond to the more luminous component which has a larger weight into the observed spectrum. 

The heliocentric RVs for the primary (V$_p$) and the secondary (V$_s$) components are listed in Table\,3, along with the dates 
of observation and the corresponding orbital phases computed with the new ephemeris given in \S 2.2. The velocities in this 
table have been corrected to the heliocentric reference system by adopting a radial velocity of 9.5 \kms for the template 
star GJ\,740. The RVs listed in Table\,3 are the weighted averages of the values obtained from the cross-correlation 
of orders \#4, \#5, \#6 and \#7 of the target spectra with the corresponding order of the standard star spectrum. The weight 
$W_i = 1/\sigma_i^2$ has been given to each measurement. The standard errors of the weighted means have been calculated on 
the basis of the errors ($\sigma_i$) in the RV values for each order according to the usual formula (e.g.\ Topping 
1972). The $\sigma_i$ values are computed by {\sc fxcor} according to the fitted peak height, as described by 
Tonry \& Davis (1979).

\begin{table}
\centering
\begin{minipage}{85mm}
\caption{Heliocentric radial velocities of NSVS\,0650. The columns give the heliocentric Julian date, the
orbital phase (according to the ephemeris in Eq.~1), the radial velocities of the two components with the 
corresponding standard deviations.
}
\label{Table 2.}
\begin{tabular}{@{}ccccccccc@{}}
\hline
HJD 2400000+ & Phase & \multicolumn{2}{c}{Star 1 }& \multicolumn{2}{c}{Star 2 } &         	& 			\\
             &       & $V_p$                      & $\sigma$                    & $V_s$   	& $\sigma$	\\
\hline
54725.4190  &0.3058 &-41.3&11.1  & 218.0  &13.6   \\
54725.4800  &0.4242 &  1.0&12.0  & 138.0  &21.3   \\
54725.5231  &0.5079 & 41.0&11.1  & --     &--     \\
54725.5675  &0.5941 & 74.0&10.1  & -50.0  &17.8   \\
54725.6102  &0.6770 &112.0& 8.4  &-108.0  &12.7   \\
54726.3640  &0.1404 &-12.0&11.5  & 172.5  &14.7   \\
54726.4069  &0.2237 &-44.4& 2.4  & 231.1  &14.2   \\
54726.4545  &0.3161 &-33.0& 9.8  & 216.6  &27.3   \\
54726.4975  &0.3996 & -9.0&11.4  & 140.1  &16.7   \\
54726.5404  &0.4829 & 24.0&14.1  & --     &--     \\
54726.5993  &0.5973 & 90.0& 9.8  & -60.0  &16.9   \\
54727.4307  &0.2114 &-34.0& 9.9  & 218.5  &17.3   \\
54727.4836  &0.3141 &-20.3& 8.8  & 225.9  &12.2   \\
54727.5161  &0.3771 &-12.0& 9.9  & 184.9  &12.2   \\
54727.5593  &0.4610 & 22.0&12.6  & --     &--     \\
54727.6005  &0.5410 & 66.0&11.8  & --     &--     \\
\hline \\
\end{tabular}
\end{minipage}
\end{table}

First we analysed the radial velocities for the initial orbital parameters. We used the orbital period held fixed 
and computed the eccentricity of the orbit, systemic velocity and semi-amplitudes of the RVs. The results of the analysis 
are as follows: $e$=0.002$\pm$0.001, i.e. formally consistent with a circular orbit, $\gamma$=44$\pm$6 \kms, 
$K_1$=77$\pm$3 and $K_2$=181$\pm$12 \kms. Using these values we estimate the projected orbital semi-major
axis and mass ratio as: $a$sin$i$=2.63$\pm$0.12 \Rsun~ and $q=\frac{M_2}{M_1}$=0.425$\pm$0.044.

\begin{table*}
\caption{Results of the V-, R-, and I-bandpass light curve analysis for NSVS\,0650. The adopted values are 
the weighted means of the values determined from the individual light curves.}
\begin{tabular}{lcccc}
\hline
Parameters &V &R  &I&Adopted  \\
\hline
$i^{o}$			               			&83.5$\pm$0.2	 & 86.5$\pm$1.3		& 81.7$\pm$0.6	 	& 83.3$\pm$0.6	\\
T$_{eff_1}$ (K)							&3\,960[Fix]	 & 3\,960[Fix]		& 3\,960[Fix]	 	& 3\,960[Fix]	\\
T$_{eff_2}$ (K)							&3\,269$\pm$51	 & 3\,401$\pm$43	& 3\,412$\pm$48	 	& 3\,365$\pm$48	\\
$\Omega_1$								&4.847$\pm$0.091 & 4.738$\pm$0.90   & 4.982$\pm$0.145	& 4.886$\pm$0.090\\
$\Omega_2$								&3.735$\pm$0.035 & 4.151$\pm$0.061	& 3.740$\pm$0.090   & 3.830$\pm$0.067\\				
r$_1$									&0.228$\pm$0.005 & 0.231$\pm$0.005	& 0.224$\pm$0.007	& 0.227$\pm$0.006\\
r$_2$									&0.176$\pm$0.003 & 0.148$\pm$0.004	& 0.171$\pm$0.006	& 0.167$\pm$0.005\\
${L_{1}}/{(L_{1}+L_{2})}$ 				&0.889$\pm$0.014 & 0.866$\pm$0.010	& 0.785$\pm$0.020	&  ------ 	\\
$\chi^2$								&1.345			 & 1.858			& 0.880				&  ------	\\				
\hline
\end{tabular}
\begin{list}{}{}
\item[$^{a}$]{\small See \S 2.1.2}
\end{list}
\end{table*}

\subsection{Light curve modeling}
As we noted in Section 2.3 the light curve of the system is considerably distorted due to light fluctuations both at 
maxima and in the deeper primary minimum.  The largest distortion with longest duration was observed on JD~24\,54746. Neither 
the amplitude nor the period or cycle of these intrinsic variations are known at this step. Therefore, we take 
all the available V-, R- and I-bandpass data for the orbital parameter analysis. The differential magnitudes of 743 in V-, 812 in R- and 
612 in I-bandpass were converted to intensities using the differential magnitudes at out-of-eclipses as $\Delta$V=1$^m$.648$\pm$0$^m$.003, 
$\Delta$R=1$^m$.198$\pm$0$^m$.001, $\Delta$I=0$^m$.575$\pm$0$^m$.002.
 
We used the most recent version of the eclipsing binary light curve modeling algorithm of Wilson \& Devinney (1971) (with updates), as 
implemented in the {\sc phoebe} code of Pr{\v s}a \& Zwitter (2005). The code needs some input parameters, which depend upon the physical 
structures of the component stars. In the light curve solution we fixed some parameters whose values can be estimated from global 
stellar properties, such as effective temperature and mass of the star. Therefore we adopted the linear limb-darkening 
coefficients from Van Hamme (1993) as 0.39 and 0.28 for the primary and secondary components, respectively; the bolometric
albedos from Lucy (1967) as 0.5, typical for a fully convective stellar envelope, the gravity brightening coefficients 
as 0.32 for the both components. The rotation of components is assumed to be synchronous with the orbital one. The mass-ratio 
of 0.425 was adopted from the semi-amplitudes of the radial velocities. We started the light curve analysis with an effective 
temperature of 3960 K for the primary star of NSVS 0650. The adjustable parameters in the light curves fitting were the orbital 
inclination, the surface potentials, the effective temperature of secondary, the luminosity of the primary.

\begin{figure*}
\includegraphics[width=12cm]{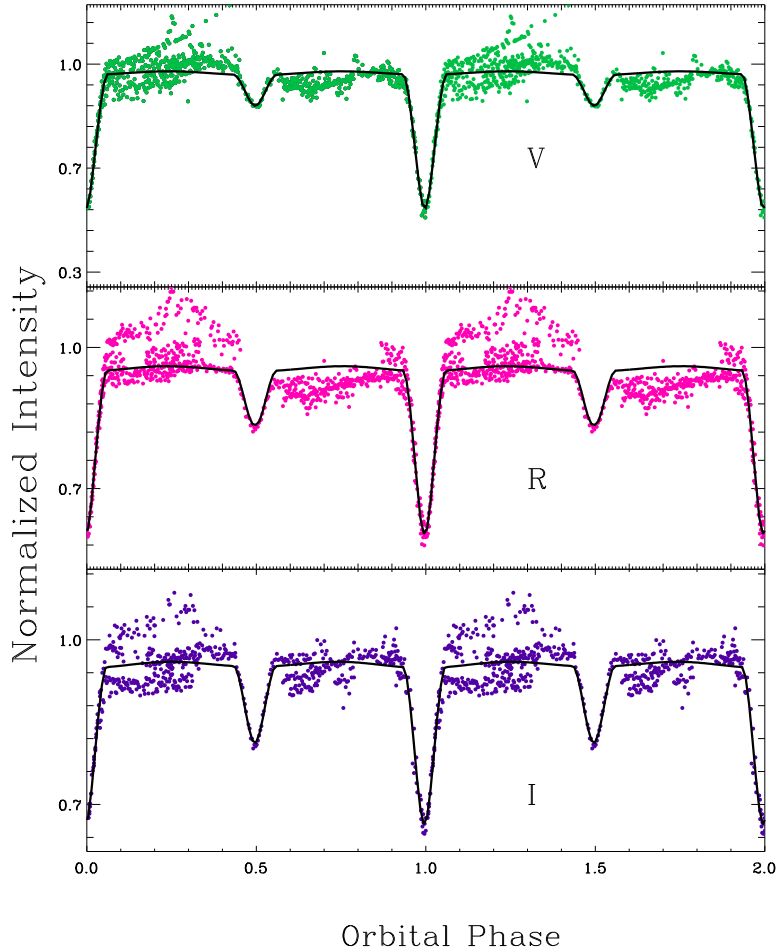}
\caption{The phase folded VRI light curves for NSVS\,0650. The best fitting solutions represented by the solid 
lines are also plotted for comparison (see text).}
\label{Figure 4.}
\end{figure*}

Using a trial-and-error method we obtained a set of parameters, which represented the observed light curves. A detached 
configuration, {\sc Mode 2}, with coupling between luminosity and temperature was chosen for solution. The iterations 
were carried out automatically until convergence, and solution was defined as the set of parameters for which the 
differential corrections were smaller than the probable errors. The orbital and stellar parameters from the 
V-, R- and I-bandpass light and radial velocity curves analysis are listed in Table 4. The uncertainties given in 
this table are taken directly from the out-put of the program. The computed light and velocity 
curves corresponding to the individual light-velocity solutions are compared with the observations in Figs. 3 and 4.

\begin{figure*}
\includegraphics[width=14cm]{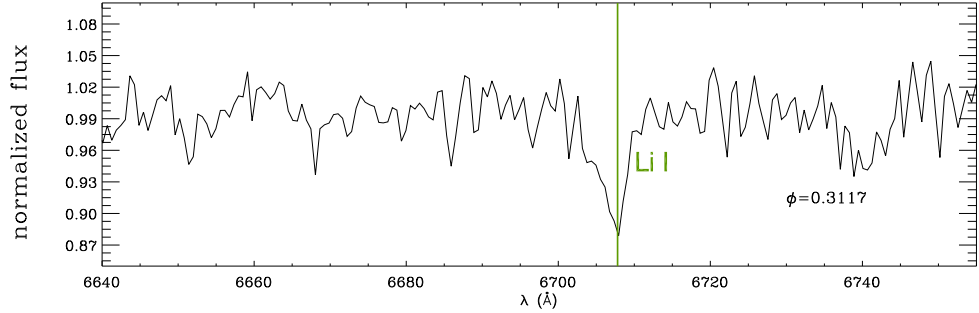}
\caption{The composite spectrum in the spectral region containing the Li I λ6708 \AA~ line observed on JD\,24\,54727.4836.}
\label{Figure 5.}
\end{figure*}

\section{The spectrum}
NSVS\,0650 has a complex spectrum over the wavelength interval from  $\sim$4100 to 8100 \AA. The spectrum is dominated
by forbidden lines and to a smaller degree, permitted emission lines of neutral metals. Strong and broad double-peaked
H$_{\alpha}$, H$_{\beta}$ and [O{\sc i}] lines are present, with the peak separation in H$_{\alpha}$ larger than the higher
{\it Balmer~lines}. The presence of the strong Li 6708 \AA~absorption line can serve as a reliable youth indicator of 
a star, as evidenced in the case of NSVS 0650. Young, low-mass pre-main-sequence stars are called T Tauri stars (TTS). They 
present the following characteristics: 1) Emission line spectra, 2) Presence of forbidden narrow-lines such as  
[O{\sc i}], [N{\sc ii}] and [Si{\sc ii}], 3) Photospheric continuum excesses (Barrado y Navascues and Martin, 2003). TTSs 
are classified into two sub-groups, the classical T Tauri stars (cTTSs) and the weak-lined T Tauri stars (wTTSs). A cTTS is surrounded by an 
optically thick disk from which it accretes material. Whereas a wTTS represents the final stage of accretion and disc-clearing 
processes (Bertout et al. 2007, Schisano et al. 2009). The equivalent width of H$_{\alpha}$ emission is used as an 
empirical criterion to distinguish between cTTS and wTTS, being smaller in the latters. Due to possible varability, no clean cut 
can be defined between the cTTS and wTTS based on the H$_{\alpha}$ emission alone. 

Spectral and photometric properties and night-to-night light variability of NSVS\,0650 indicate that the active star in the 
system resembles many characteristics of the TTSs as given above and discussed by Alcala et al. (1993), Covino et al. 
(1996), and Alencar \& Basri (2000). As it is known  the optical emission lines  are definite characteristics of  the many 
late type, main-sequence systems, including NSVS\,0650. Another fundamental characteristic of TTSs is the variations 
of H-line profiles (Ferro \& Giridhar 2003). NSVS\,0650 is composed of low-mass stars which cover most of the properties of 
the T Tauri stars. 

\begin{figure*}
\includegraphics[width=11cm]{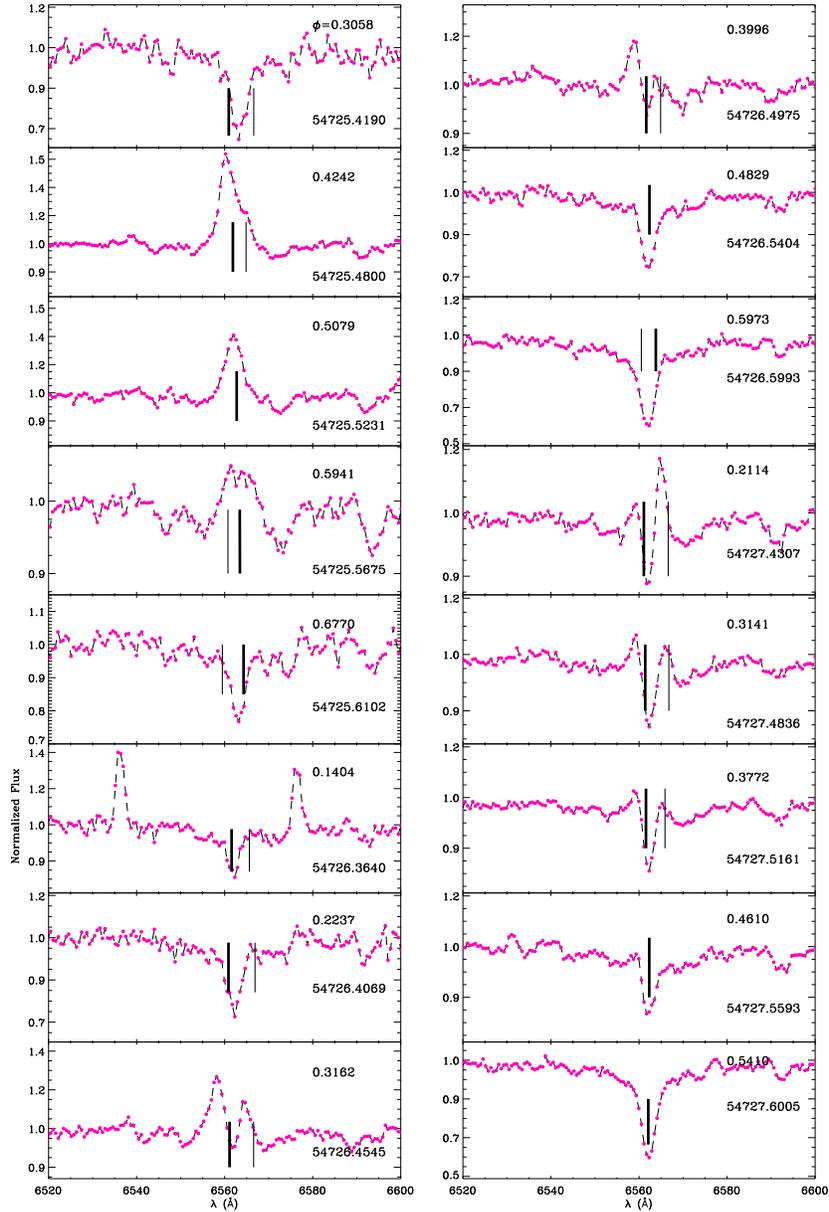}
\caption{Variation of the H$_{\alpha}$ line profiles of NSVS\,0650. The normalized spectrum at the H$_{\alpha}$ ordered with 
the orbital phase. The vertical thick and thin lines show the rest wavelenghts corresponding to the primary and secondary
component photospheres, respectively.}
\label{Figure 6.}
\end{figure*}

\begin{figure*}
\includegraphics[width=10cm]{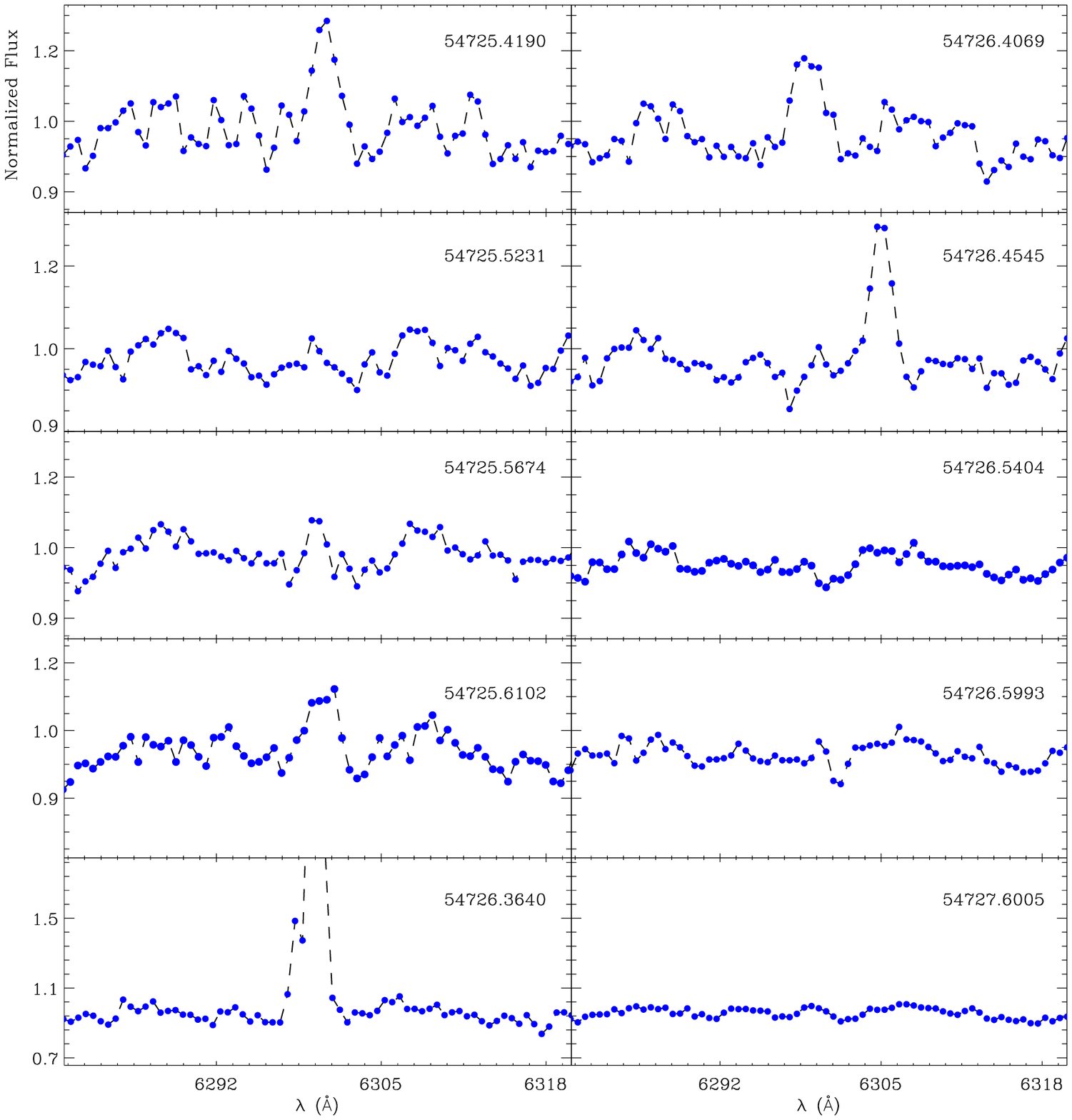}
\caption{Variation of the forbidden emission line profiles of the [O{\sc i}] at 6300 \AA. }
\label{Figure 7.}
\end{figure*} 

\subsection{Line profiles}
The most conspicuous line with dramatic profile variations in the system's spectrum appears to be the  H$_{\alpha}$. The 
H$_{\alpha}$ line is the most prominent feature in the spectra of TTSs. The presence of the Li absorption line 
at 6708 \AA~ (see Fig. 5, for an example) and weak H$_{\alpha}$ in emission leads us to classify the star as weak-lined
T Tauri star. In Figure 6 we display the H$_{\alpha}$ line region observed at various orbital 
phases in three consecutive nights. Each spectrum has been normalized to the continuum. Julian date and the orbital phase 
for each observation are given in each panel. On JD\,2454725 the H$_{\alpha}$ line appears to be a single, 
shallow absorption, i.e., filled-in by emission, at orbital phase of about 0.3058. At orbital phases of 0.4242, 0.5079 
and 0.5939 the same line becomes single, emission above the continuum and at phase of 0.6770 it turns to be an absorption 
again. On JD\,2454726, the following night, the H$_{\alpha}$ line is seen as single absorption at phases of 0.1404 and 
0.2237, whereas double-peaked emission profiles at phases of 0.3162 and 0.3996, but it turns to absorption in a short 
time interval at phases of 0.4829 and 0.5973. The H$_{\alpha}$ emission line profile at orbital phase of 0.3996 has an 
unexpected shape because it is very resemblance of {\it inverse P Cygni profile}, most similar to UX\,Tau A (see Reipurth 
et al. 1996). The dramatic changes in the shape of the H$_{\alpha}$ line, collected on JD\,2454727, are clearly seen in 
the last five panels of Fig. 6. The H$_{\alpha}$ line in the spectra of the NSVS\,0650 taken at phases of about 0.2114, 0.3141 
and 0.3772 displays blue-shifted absorption, similar to wTTS GG\,Tau (Folha \& Emerson, 2001). It turns to be single 
absorption at orbital phases of 0.4610 and 0.5410.  

The higher Balmer series,  H$_{\beta}$  and H$_{\gamma}$ lines of NSVS 0650 generally appear to be in emission at all 
orbital phases. Again, dramatic line profile changes are evident. Inverse P Cygni profiles are also visible at 
some orbital phases. 

The existence of blue-shifted absorption components in the Balmer lines of TTSs' spectra was first noted by Herbig (1962), 
who suggested that these absorption components are evidence for strong stellar winds. On the other hand Walker (1972) 
drew attention to the wTTSs which have red-shifted absorption in the higher-order Balmer lines. These inverse P Cygni 
profiles have generally been interpreted in terms of material accreting onto the young stars. The optical observations of 
unidentified Einstein Observatory X-ray satellite sources led to the the discovery of many TTSs with weak H$_{\alpha}$ and 
IR excess emission (Strom et al. 1990). The wTTSs have also dark spots as in the case of cTTSs but stronger X-ray 
in emission than cTTSs. They have also shallow or no disks. If a wTTS has still a disk some winds are blown away from this 
disk. Most TTSs are members of close binaries wich may be born without a disk or have a short-lived disk (Neuhauser, 
1997). Three types of binaries including TTS without disks, with circumstellar disks and with circumbinary disks exist.

\subsection{Forbidden lines}
One of the most important characteristics in the spectra of the cTTSs is the presence of forbidden emission lines. The forbidden 
neutral oxygen lines are not seen in the spectra of wTTSs.  In the spectra of NSVS 0650 we observed forbidden [O{\sc i}] emission line at 6300 \AA. 
Figure 7 displays the [O{\sc i}] emission line profiles at various orbital  phases. [O{\sc i}] emission line shows single peaks, but the line centroid 
is shifted to the blue. The strength of emission in [O{\sc i}] 6300 \AA~ is highly variable and seems 
to correlate with the orbital phase. This forbidden emission line appears to slightly stronger at the first quadrature 
than at the second one. In the optical spectrum of cTTSs the forbidden emission lines are dominated. These lines are usually 
patterns of shocked low-density regions of young stars (Fernandez \& Cameron 2001). These shocks can be produced by the 
outflowing materials, winds, and/or jets. Strong H$_{\alpha}$ and [O{\sc i}] emission 
lines in the optical spectra of NSVS 0650 are indicative of ongoing accretion. 
The strength of H$_{\beta}$ and its equivalent width (EW) seems to correlate well with that of  H$_{\alpha}$, as shown in Fig. 
8. However the EWs of [O{\sc i}] do not correlate well with those of the H$_{\alpha}$, as is seen in the upper panel of Fig. 8. It 
appears that as if there is an anti-correlation between [O{\sc i}] and H$_{\alpha}$. The range of variation in the EWs of 
H$_{\alpha}$ is between about 2 \AA~ and about -3 \AA.  Whereas the EWs of [O{\sc i}] vary from 0 to about -1 \AA. However, we 
observed the most strong emission in [O{\sc i}] on JD 2454726.3640 at orbital phase of 0.1404 with an EW of -2.2 \AA. We
also measured the average EW of Li{\sc i} as 0.3$\pm$0.2 \AA.

\begin{figure}
\includegraphics[width=7.5cm]{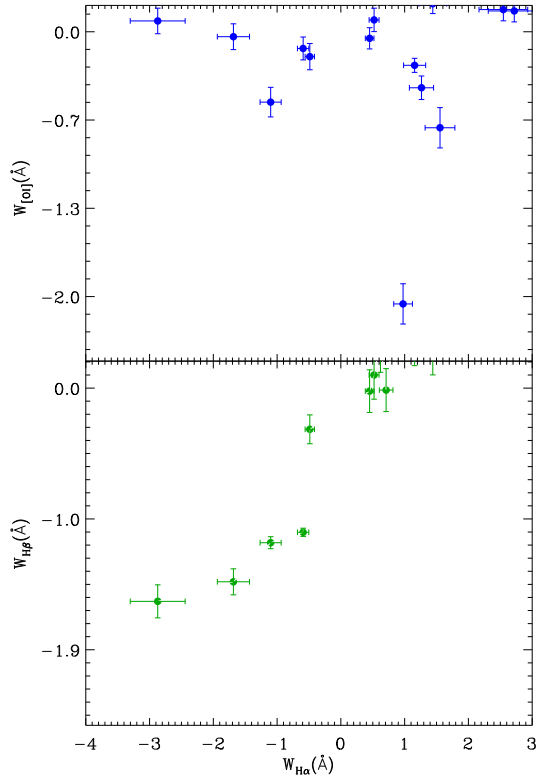}
\caption{Correlations between equivalent witdh of the most prominent emission lines measured in the spectra.}
\label{Figure 8.}
\end{figure}

 \begin{table}
 \setlength{\tabcolsep}{2.5pt} 
  \caption{Fundamental parameters of the system.}
  \label{parameters}
  \begin{tabular}{lcc}
  \hline
   Parameter 						& Primary	&	Secondary	\\
   \hline
   Mass (M$_{\odot}$) 				& 0.656$\pm$0.086 & 0.279$\pm$0.045	\\
   Radius (R$_{\odot}$) 			& 0.600$\pm$0.030 & 0.442$\pm$0.024	\\
   $\log~g$ ($cgs$) 				& 4.699$\pm$0.032 & 4.594$\pm$0.047	\\
   $T_{eff}$ (K)					& 3\,960$\pm$80	& 3\,365$\pm$80 	\\
   $(vsin~i)_{calc.}$ (km s$^{-1}$)	& 59$\pm$3 & 43$\pm$2	 	      	\\
   $\log~(L/L_{\odot})$				&-1.097$\pm$0.057	& -1.647$\pm$0.062  	\\
   $d$ (pc)							& \multicolumn{2}{c}{111$\pm$9}		\\
   $J$, $H$, $K_s$ (mag)$^{*}$		& \multicolumn{2}{c}{10.918$\pm$0.032, 10.267$\pm$0.028, 10.092$\pm$0.025}	\\
$\mu_\alpha$, $\mu_\delta$(mas yr$^{-1}$)$^{**}$ & \multicolumn{2}{c}{23.70$\pm$6.10, 17.50$\pm$6.10} \\
$U, V, W$ (km s$^{-1}$)  & \multicolumn{2}{c}{-40$\pm$5, +19$\pm$4, -14$\pm$3}\\ 
\hline  
  \end{tabular}
\medskip\\
{\rm *{\em 2MASS} All-Sky Point Source Catalogue (Cutri et al. 2003)} \\ 
{\rm **NOMAD Catalog (Zacharias 2005)} \\ 
\end{table}

\section{Discussion}
One of the goals of the present study is to derive the physical parameters of the low-mass stars in the eclipsing binary 
systems. As it is known eclipsing binaries are the most suitable laboratories for determining the fundamental properties 
of the stars and thus for testing the predictions of theoretical models. For this reason we started optical photometric 
and spectroscopic observations of some selected low-mass stars. We obtained multi-band light curves and spectra with a wide 
wavelength range. We analyzed the V-, R-, and I-bandpass light curves and the radial velocities separately using the modern 
codes. Then, we combined the photometric and spectroscopic solutions and derived the absolute parameters of the component 
stars. The standard deviations of the parameters have been determined by JKTABSDIM\footnote{This can be obtained from http://http://www.astro.keele.ac.uk/$\sim$jkt/codes.html} code, which calculates distance and other physical parameters 
using several different sources of bolometric corrections (Southworth et al. 2005). The best fitting parameters are 
listed in Table 5 together with their formal standard deviations.

The luminosity and absolute bolometric magnitudes M$_{bol}$ of the stars were computed from their effective 
temperatures and their radii. Since low-mass stars radiates more energy at the longer wavelengths we used $VRIJHK$ 
magnitudes given by Coughlin \& Shaw (2007). Applying $BVRIJHKL$ magnitudes-T$_{eff}$ relations given by 
Girardi (2002) we calculated the distance to NSVS\,0650 as $d$=111$\pm$9 pc. Estimating distances to low-mass 
stars are strongly depended on the bolometric corrections. If we adopt the bolometric corrections given by 
Siess, Forrestini \& Dougados (1997) the distance to NSVS 0650 reduces to about $d$=86$\pm$4 pc. The mean light contribution of 
the secondary star $L_2$/($L_1$+$L_2$)=0.22 obtained directly from the I-bandpass light curve analysis is in a good 
agreement with that estimated from the bolometric luminosities as 0.22. This result indicates that the light contribution
of the less massive component is very small, indicating its effect on the color at outsite eclipse is very limited.

Locations of the primary and secondary components on the theoretical mass-radius and T$_{eff}$ - log L/L$_{\odot}$
diagrams are shown in Fig. 9. The mass tracks and isochrones are adopted from Siess, Forrestini \& Dougados (1997) and 
Siess (2000). Since the stars appear to be in pre-main sequence evolution we adopted Z=0.03. These mass tracks are 
very close to those obtained for solar abundance.
The radius of more massive primary component is in agreement with that zero-age main sequence star having 
the same mass. However, the secondary is about 35 \% larger than the main-sequence counterpart. This result confirm the hypothesis 
proposed by Chabrier, Gallardo \& Baraffe (2007) for the larger radius of the low-mass convective stars. 
The existence of Li 6708 \AA~ absorption 
line in the spectra and comparison the absolute dimensions of the components with the evolutionary tracks may be taken 
as an indicator of the pre-main-sequence stars. The components of NSVS\,0650 lie on the isochrones between 15-30 Myr, still 
in contracting phase toward the main-sequence. If we use the isochrones plotted M$_v$ versus $B-V$  we estimate an age of about 
10-15 Myr. This difference arises from the bolometric corretions given by Siess, Forrestini \& Dougados (1997).  We used the 
color-temperature calibrations given for the main-sequence stars for estimating the effective temperature of the more massive 
primary component. If we use the color-temperature relation given for luminosity class IV stars (de Jager and Nieuwenhuijzen, 1987) we find even 
smaller effective temperature af about 3\,700 K for the primary component. The 
difference of about 250 K in the effective temperature of the primary star shifts its location to the lower-right in 
the HR diagram which corresponds to a smaller age.  We estimate an age of about 50 Myr using the pre-main-sequence models 
given by D'Antona \& Mazzetelli (1997), and 63 Myr by Baraffe et al. (1998) slightly larger than that given by Siess et al. 
model. The evolutionary models indicate that a star with a mass of 0.66 \Msun~ takes about 100 Myr to contract and reach 
its normal main-sequence radius.

\begin{figure*}
\includegraphics[width=15cm]{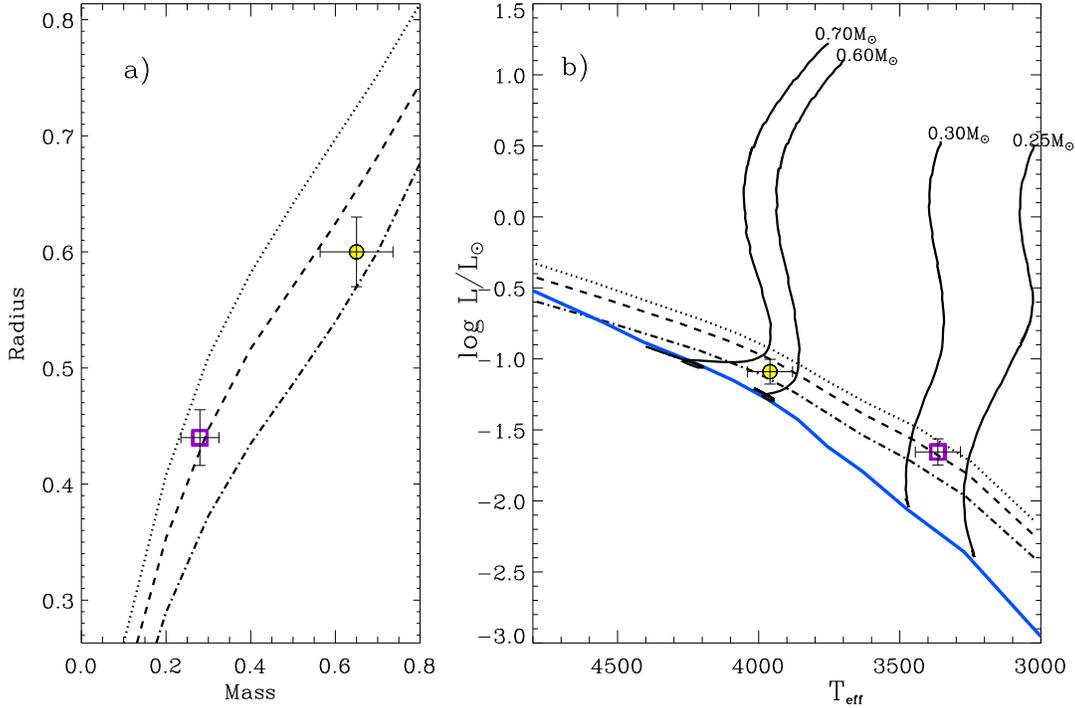}
\caption{Comparison between stellar models and the absolute dimensions of NSVS\,0650 in the mass-radius (a) 
and $T_{eff}$ - $\log~(L/L_{\odot})$ (b) planes. The mass-radius relations in panel (a) were derived using the stellar models
of Siess et al. (2000) for Z=0.03 with an age of 15 (dotted), 20 (dashed) and 30 Myr (dot-dashed). Panel (b) shows 
the locations of the components in the HR diagram. Evolutionary tracks for the masses of 0.25, 0.30, 0.60 and 0.70
\Msun~ are shown for comparison. The diagonal lines from left to right indicate isochrones with an age of 
15 (dotted), 20 (dashed) and 30 Myr (dot-dashed) and zero-age main-sequence (continuous line). The filled-circle and 
square indicate the primary and secondary components of NSVS\,0650, respectively.}
\label{Figure 9.}
\end{figure*} 

The Li{\sc i} 6708 \AA~ line is often used an age indicator. In the spectra of NSVS\,0650 the Li{\sc i} line is clearly 
seen. Moreover, in the optical spectrum of the system, we observed also H$_{\alpha}$ and H$_{\beta}$ lines as in 
emission. In addition, strong emission of the [O{\sc i}] forbidden line is visible. These features are signs of 
cTTS but, in contrary, the measured EW values point a wTTS. The primary component of NSVS\,0650 appears in the region 
of $Li-poor$ stars located on the HR diagram (see Fig. 8 in Sestito, Palla \& Randich, 2008). The measured EW of 
0.3 \AA~ for Li{\sc i} is in agreement with this classification. High-resolution spectra 
are urgently required to confirm our finding and to derive which sub-group, cTTS or wTTS, it belongs.

\section{Acknowledgements}
The authors acknowledge generous allotments of observing time at TUBITAK National Observatory (TUG) of Turkey. We also 
wish to thank the Turkish Scientific and Technical Research Council for supporting this work through grant Nr. 108T210 
and  EB{\.I}LTEM Ege University Science Foundation Project No:08/B\.{I}L/0.27 . We have use of 2MASS USNO and NOMAD Catalogs
as well as the Simbad, Visir, and ADS databases.  The anonymous referee is gratefully acknowledged for useful and constructive 
suggestions.

\end{document}